# Theory of Ergodic Quantum Processes

Ramis Movassagh[1,*] and Jeffrey Schenker[2,†]

[1]*IBM Quantum Research, MIT-IBM AI lab, Cambridge, Massachusetts 02139, USA*
[2]*Department of Mathematics, Michigan State University, East Lansing, Michigan 48824, USA*

(Received 11 September 2020; revised 4 March 2021; accepted 23 July 2021; published 1 October 2021)

The generic behavior of quantum systems has long been of theoretical and practical interest. Any quantum process is represented by a sequence of quantum channels. We consider general ergodic sequences of stochastic channels with arbitrary correlations and non-negligible decoherence. Ergodicity includes and vastly generalizes random independence. We obtain a theorem which shows that the composition of such a sequence of channels converges exponentially fast to a replacement (rank-one) channel. Using this theorem, we derive the limiting behavior of translation-invariant channels and stochastically independent random channels. We then use our formalism to describe the thermodynamic limit of ergodic matrix product states. We derive formulas for the expectation value of a local observable and prove that the two-point correlations of local observables decay exponentially. We then analytically compute the entanglement spectrum across any cut, by which the bipartite entanglement entropy (i.e., Rényi or von Neumann) across an arbitrary cut can be computed exactly. Other physical implications of our results are that most Floquet phases of matter are metastable and that noisy random circuits in the large depth limit will be trivial as far as their quantum entanglement is concerned. To obtain these results, we bridge quantum information theory to dynamical systems and random matrix theory.

## I. OVERVIEW AND SUMMARY OF THE RESULTS

How do generic quantum systems behave? The physical change of a quantum system over a unit of time is modeled by the application of a quantum channel. Quantum channels are the most general formulation of physical (quantum) processes such as various steps of a quantum computation, effects of noise and errors on the state, and measurements [1]. Any particular evolution of a quantum system can be obtained by the application of a sequence of suitably chosen quantum channels.

The generic (average) behavior of quantum systems has long been of theoretical and practical interest. Random channels appear in a wide variety of applications, from quantum chaos [2] to holographic dualities in theories of quantum gravity [3] to operator dynamics [4]. Similarly, random local circuits are currently intensely studied for their $k$-design properties [5], potential to demonstrate quantum supremacy [6,7], and understanding operator spread and entanglement growth [8]. Recently, motivated by the demonstration of quantum supremacy, random circuits have become central candidates to show hardness in near-term quantum computing. For example, Google recently demonstrated a 53-qubit experimental demonstration of the hardness of sampling from the output of random circuits [9].

In nature, whenever one studies a quantum system, leakage of information to the environment is at play. A key challenge is to address the effects of decoherence in natural settings or account for it in the lab especially now that quantum error correction schemes have not been realized. To model decoherence on the output of a quantum circuit, scenarios have been considered in which after every gate a measurement is performed with some probability [10]. Here, we take a different perspective in which decoherence is directly included in the channels describing the process.

In this paper, we consider a sequence of channels as a trajectory of a dynamical system. Dynamical systems theory is a rich field with seminal results such as the Furstenberg-Kesten theorem [11] and the multiplicative ergodic theorem of Oseledec [12]. By bridging quantum information to the theory of dynamical systems, we address the behavior of quantum channels in a very general setting and in the presence of decoherence. We take the underlying dynamical system and the sequence of channels to be ergodic. Ergodicity allows for arbitrarily long-range correlations among the channels, including and vastly generalizing the extreme cases of independently and identically distributed (IID), as well as (time)-translational-invariant, channels (Fig. 1). Below, we give a more formal definition.

*ramis@us.ibm.com
†jeffrey@math.msu.edu

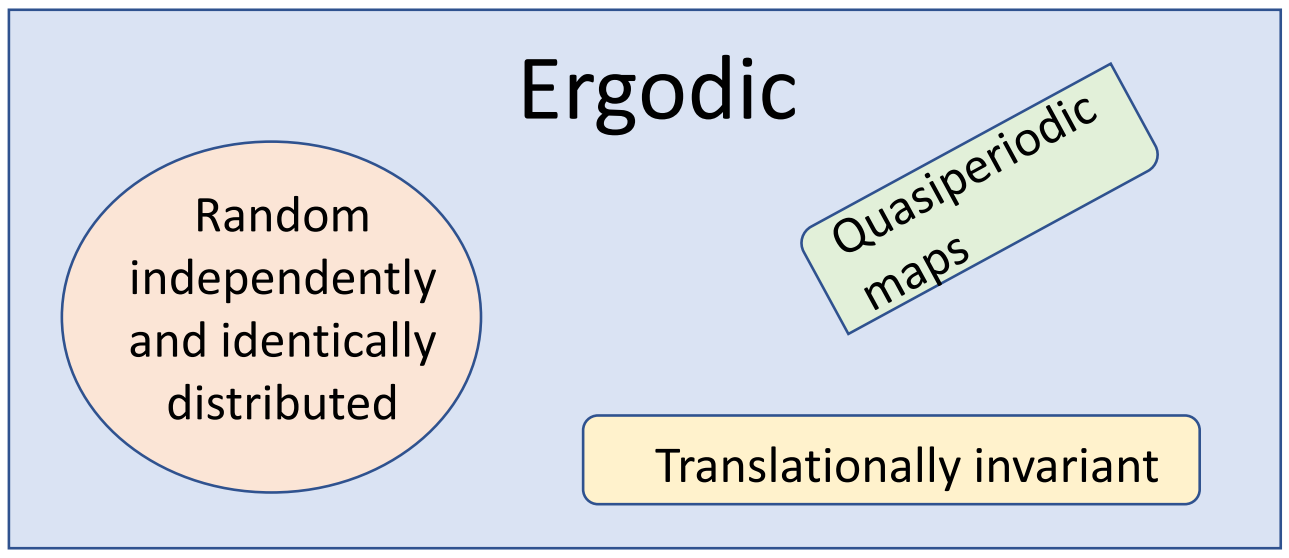

FIG. 1. Venn diagram for distributions of possible ergodic sequences of channels. Among the many subsets, three subsets are shown and substantially enlarged for readability.

We first present a general theorem for an ergodic sequence of quantum channels (Theorem 1), which shows that the composition of such channels converges exponentially fast to a stochastic sequence of rank-one channels. A corollary of this result is the well-known convergence in the translation-invariant case to a fixed rank-one channel. We then apply our theorem in a natural setting in which Kraus operators [$B_m^{i_m}$ in Eq. (2)] in each channel are IID random Haar isometries, where by Haar isometry we mean a section cut out of a unitary matrix that is uniformly drawn from the space of all possible unitary matrices. We analyze the asymptotics with respect to the environment or the system or both tending to infinity and prove universal Gaussianity in these limits.

In the past, "ergodic" quantum channels were considered, which, to the best of our knowledge, assumed special subsets of possibilities considered herein. For example, a channel is chosen at random from some ensemble and then repeatedly applied [13]. In other work, time dynamics are analyzed for a quantum system with repeated independently chosen random interactions with an environment [14], and convergence results are obtained for the expected behavior of a sequence of IID channels and their applications [15]. Other instances studied include certain random independent channels and their compositions (e.g., from a finite set of random isometries) [16,17]. See Ref. [18] for a review. Our work considers a general ergodic sequence and, we believe, serves as a vast generalization of the prior work. We emphasize that this work allows for correlated randomness or even pseudorandomness [19] generated by quasiperiodic dynamics.

Interestingly, the formalism of quantum channels is suited for the calculation of physically relevant quantities such as expectation values of observables and correlation functions of low-dimensional quantum systems. Such quantum systems are best described by the density matrix renormalization group (DMRG), whose natural representations are matrix product states (MPS).

Theorem 1 allows us to analyze the thermodynamic limit of ergodic MPS. We derive formulas for the expectations of a general observable and for the bipartite entanglement entropy across an arbitrary cut. The second main result of our paper is Theorem 2 on the exponential decay of two-point functions for ergodic MPS.

Any MPS is a ground state of a nontrivial Hamiltonian. Furthermore, MPS are the natural representation of states obtained from DMRG.

The tensors in the MPS representation of the ground state of any interacting Hamiltonian (any nontrivial Hamiltonian) *will be correlated*. This result is so *even if the underlying Hamiltonian has random independent local terms*. In particular, a state obtained from a DMRG procedure, say, the ground state of a 1D spin glass, will be correlated yet fall within the class of ergodic MPS.

This work, to the best of our knowledge, is the first to shed light on the structure of correlations for a large general class of correlated MPS that one obtains using procedures such as DMRG. This class includes MPS with IID local tensors. However, as the above remarks show, IID local tensors are mostly of theoretical interest, as correlations inevitably arise in any realistic interacting system even if the underlying Hamiltonian is fully disordered.

More broadly, it is important to explore the manifold of non-translation-invariant MPS. The present work provides a well-motivated path for further study of this line of research.

Despite various efforts over about the past two decades, previous results in the same vein as this paper were obtained only for very special classes of aforementioned quantum processes, with limited applicability to physically realistic settings.

We achieve our results by bridging the mathematical fields of dynamical systems and random matrix theory to quantum information science. Roughly speaking, Theorem 1 can be seen as the "quantization" of classical results on the convergence of Markov chains. The proof is based on generalizations of the Perron-Frobenius theorem and of the multiplicative ergodic theorem. The analysis of Haar channels is enabled by the use of the Weingarten calculus, with Gaussianity of the fluctuations established by verifying Wick's theorem. Full proofs and derivations are given in the appendixes and in Ref. [20].

In Sec. II A, we detail further consequences of our work for physics of matter, quantum information theory, and MPS.

## II. ERGODIC THEOREM FOR COMPOSITION OF QUANTUM CHANNELS

A quantum channel is a completely positive and trace-preserving linear map $\phi$ on the space of $D \times D$ matrices:

$$\phi(\rho) = \sum_{i=1}^{d} B^i \rho B^{i\dagger}. \tag{1}$$

The total change to a quantum state is obtained by the composition of suitably chosen such maps on the initial state (i.e., density matrix $\rho$):

$$\phi_n \circ \cdots \circ \phi_0(\rho) = \sum_{i_n,\ldots,i_0} B_n^{i_n} \ldots B_0^{i_0} \rho B_0^{i_0\dagger} \ldots B_n^{i_n\dagger}. \tag{2}$$

We consider sequences $(\phi_0, \phi_1, \ldots)$ of channels drawn from an ensemble which is ergodic with respect to the time shift $(\phi_0, \phi_1, \ldots) \overset{T}{\mapsto} (\phi_1, \phi_2, \ldots)$. In statistical physics, originating in the fundamental work of Boltzmann, the ergodic hypothesis is traditionally understood as the equivalence of time averages and ensemble averages [21]. Later, the ergodic hypothesis was made more precise in mathematical physics culminating in the works of von Neumann and Birkhoff [22]. In those works, it is recognized that the ergodic hypothesis of Boltzmann is a consequence of the ergodic nature of the underlying dynamics. Formally, a map $T$ on an ensemble is ergodic if it preserves probabilities and, starting from a typical point, the dynamics $T$ generates covers the phase space with probability one (we recall the precise mathematical definition in Appendix A).

In general, there may be selection rules separating distinct sectors of the Hilbert space in which the quantum state $\rho$ resides that do not mix under the evolution (i.e., application of channels). Furthermore, there may be a subspace that is transient for the evolution, such as the space of excited states of a system for which the long-time evolution is described by decay to a ground state. The long-time behavior of the system can be understood by restricting attention to a single selection sector orthogonal to the transient subspace. Without loss of generality, we can assume that our system (i) has no selection rules—there are no (proper) subspaces invariant under the evolution, because if there were such a subspace, we could simply study each one separately—and (ii) has no transient subspace, because if it does, then the systems eventually leaves it with a probability arbitrary close to one.

*Assumptions.*—Mathematically, the only assumption we need is that with probability one there exists an integer $N_0 > 0$ such that for all $N > N_0$, the map $\phi_N \circ \phi_{N-1} \circ \cdots \circ \phi_0$ is strictly positive. This is implied by the following more easily verifiable assumptions.

(1) For some $n_0$,

$$\mathrm{Prob}[\phi_{n_0} \circ \cdots \circ \phi_1 \circ \phi_0 \text{ is strictly positive}] > 0.$$

(2) With probability one, if, for some observable $M \geq 0$, we have $\mathrm{tr}[M\phi_0(\rho)] = 0$ for all $\rho$, then $M \equiv 0$.

Assumption 1 is a generalization to ergodic quantum processes of the notion of irreducibility for a finite Markov chains, for which the transition matrices can be analyzed by the Perron-Frobenius theorem.

In the absence of selection rules (or upon reduction to an appropriate selection sector), these assumptions are expected to hold for any system interacting with an environment at positive temperature (but would not hold at zero temperature, where excited states are transient). More generally, they should hold provided there is non-negligible decoherence, leading to the possibility of a transition between any two states of the system. Note that the second assumption can alternatively be expressed as the following condition on the dual channel: For a non-negative observable $M$, if $\phi_0^*(M) = 0$, then $M = 0$.

For each $m, n \in \mathbb{Z}$ with $m < n$, let the quantum channel $\Psi_{n,m}$ be $\Psi_{n,m} \equiv \phi_n \circ \cdots \circ \phi_m$. Our main result is the following theorem.

*Theorem 1.*—There exists $0 < \mu < 1$ and an ergodic sequence of $D \times D$ density matrices $(\ldots, Z_m, Z_{m+1}, \ldots, Z_n, Z_{n+1}, \ldots)$ such that, given $x \in \mathbb{Z}$, the following holds:

$$\|\Psi_{n,m}(\rho) - Z_n\| \leq C_x \mu^{n-m} \tag{3}$$

for all $m \leq x$, $n \geq x$, and any density matrix $\rho \in \mathbb{C}^{D\times D}$, where $C_x < \infty$ almost surely. Furthermore, the matrices $Z_j$ satisfy the shift equations

$$Z_j = \phi_j(Z_{j-1}). \tag{4}$$

The matrices $Z_j$ are analogous to the microstates of a quantum system at equilibrium, and the recursion given by Eq. (4) is the transition from one microstate to the next dictated by the ergodic quantum process. The theorem above proves that any initial state $\rho$, which may be far from equilibrium, approaches equilibrium exponentially fast in the following sense. The dynamical trajectory of any initial state, after long enough evolution with respect to the application of the ergodic sequence of the quantum channels, becomes exponentially close to the trajectory of the equilibrium microstates.

The parameter $\mu$ is a characteristic of a given ergodic quantum process; it depends on the ensemble describing of the process but is nonrandom in that it does not depend on the specific realization of a sequence of channels from the ensemble. The logarithm $\lambda = -\log\mu$ is a Lyapunov exponent giving the inverse relaxation time of the process. The prefactors $C_x$ appearing in Eq. (3) are, on the other hand, random (dependent on the specific realization) and as indicated depend on the origin of time denoted by $x$.

A key point is that for $n - m$ sufficiently large $\Psi_{n,m}$ is exponentially close to the replacement (rank-one) channel $\Psi_{n,-\infty}(\rho) = Z_n \mathrm{tr}[\rho] = Z_n$. This result is largely a consequence of non-negligible decoherence (assumption 1 above).

An important feature of this convergence is that, although the *distribution* of $Z_n$ is fixed in time, the matrices

$Z_n$ are random objects that fluctuate with respect to the time step $n$.

By ergodicity, the distribution of $Z_n$ can be determined from the frequency of occurrences in the trajectory $Z_n, Z_{n+1}, \ldots$. A trivial example is furnished by the IID rank-one channels $\phi_j$:

$$\phi_j(\rho) = \begin{cases} \rho_0 \operatorname{tr}[\rho] & \text{with probability } 1/2, \\ \rho_1 \operatorname{tr}[\rho] & \text{with probability } 1/2, \end{cases}$$

with two distinct density matrices $\rho_0$ and $\rho_1$. In this case, each $Z_n$ has the distribution

$$\mathrm{Prob}[Z_n = \rho_0] = \mathrm{Prob}[Z_n = \rho_1] = \frac{1}{2},$$

and the individual trajectories fluctuate randomly between the two values $\rho_0$ and $\rho_1$, each occurring with frequency 1/2.

To fully characterize the statistical equilibrium described by the process $\ldots, Z_j, \ldots$, we need to solve the recursion in Eq. (4). This solution depends on the specification of the channels. To demonstrate the theory, we consider two natural extreme cases of an ergodic sequence of channels. First, we take the translationally invariant channels in which every channel in the sequence is equal. We then consider the other extreme in which the $B_k^{i_k}$ in Eq. (2) are blocks of IID Haar unitaries.

### A. Physical consequences

We have quantifiable bounds and theorems that, to the best of our knowledge, for the first time apply to correlated quantum processes and to the ground states of interacting quantum matter (see Secs. I and IV). Any realistic physical process inevitably has temporal correlations, which is so even if the underlying process is Markovian, for which two subsequent times are correlated. We emphasize that correlated channels naturally arise in the study of nontrivial systems, such as interacting quantum many-body systems (see Secs. I and IV). Therefore, an IID assumption is of purely theoretical interest. This work, however, applies to a vast class of physically realistic quantum processes.

Three physical corollaries to our theorem are as follows.

(a) Nonequilibrium phases of matter that are engineered by time-periodic driven Hamiltonians, such as in Floquet systems [23–25], *must* be metastable when interactions with an environment in positive temperature are non-negligible (equivalent of the assumption above). Therefore, generically any experimentally viable Floquet phase can only be metastable.

(b) These channels become trivial as far as their information content is concerned. In fact, since the channels are asymptotically replacement channels, the process cannot even convey classical information. This constraint can be seen intuitively by the fact that a unique fixed point is reached irrespective of the input initial quantum state. We emphasize that channels may be highly correlated; this possibility holds even in the time-translation-invariant case. For example, in near-term quantum computing with noisy random circuits [9,26,27], in the limit of large depth, all initial memory of the state is lost and a unique final state results.

(c) A constant gap implies an exponential decay of correlations [28,29] and in one dimension an area law for entanglement entropy [30]. Later, Brandão and Horodecki [31] proved that in one dimension exponential decay of correlations implies an area law. Below in Sec. IV, we prove a partial converse, which says that finitely correlated states with an ergodic MPS representation have an exponential decay of correlation.

## III. APPLICATIONS

In the translationally invariant case, the channels are the same, $\phi_j = \phi_0$ for all $j$, and assumption 1 reduces to the requirement that $\phi_0^{n_0}$ be strictly positive for some $n_0$. The resulting sequence $Z_j$ given by Theorem 2 is also translation invariant, $Z_j = Z_0$. Indeed, in this case, one may easily show that $Z_0$ is the unique eigenmatrix of $\phi_0$ with eigenvalue 1, and all other eigenvalues of $\phi_0$ have modulus smaller than $1-\mu$. Equation (3) shows the exponentially fast convergence of the density matrix to the state $Z_0$ under repeated application of the channel $\phi_0$. That is, starting from any density matrix, repeated application of the channel derives the system exponentially fast toward the equilibrium state $Z_0$.

In general, $Z_j$ is not an eigenmatrix of $\phi_j$; rather, these matrices obey Eq. (4). Since $Z_{j-1}$ and $Z_j$ are drawn from the same probability distribution, Eq. (4) implies a fixed point equation for the distribution of $Z_j$. For general ergodic channels, the precise form of the fixed point equation depends on the correlations between the channels. For IID channels, the fixed point equation takes a particularly simple form whose solution can be determined by the method of moments (see Appendix B).

To illustrate this implication in a natural example, consider the other extreme and take the sequence of channels $\phi_j(\rho) = \operatorname{tr}_r[U_j \rho \otimes Q_r U_j^\dagger]$, where $Q_r$ is a pure state (rank-one projection) in $\mathbb{C}^{r\times r}$, which we think of as the environment, and $\rho \in \mathbb{C}^{D\times D}$ is the initial state of the system. Operators $U_j$ are a sequence of independent Haar distributed $Dr \times Dr$ dimensional unitaries. Random Haar channels are the archetypes of generic quantum channels. Properties of a single random channel of this type are considered in Ref. [32]. Here, we consider an ergodic process obtained from an IID sequence of such channels.

We show that the average of the equilibrium matrices $Z_j$ is the totally mixed state (see Appendix B)

$$\mathbb{E}\{Z_j\} = D^{-1} I_D. \tag{5}$$

Furthermore, using the Weingarten calculus [16,33], we show that fluctuations around the average are given by

$$Z_j = D^{-1} I_D + (1 + rD^2)^{-1/2} W_j, \tag{6}$$

where $\mathrm{tr}[W_j] = 0$ and. in the limit that $D \to \infty$ or $r \to \infty$ (or both), $W_j$ has a Gaussian distribution on the space of $D \times D$ matrices:

$$\mathrm{Prob}[W_j = W] = \frac{1}{\Xi_D} e^{-D/2\mathrm{tr}[W^2]} \delta(\mathrm{tr}[W]) dW, \tag{7}$$

where $\Xi_D$ is a $D$-dependent normalization (see Appendix B). As a result, for large $D$ the eigenvalues of $Z_j$ are distributed according to Wigner's semicircle law:

$$\mathrm{tr} f(Z_j) \approx \frac{2D}{\pi} \int_{-1}^{1} dt \sqrt{1 - t^2} f\left(\frac{1}{D} + Lt\right)$$

with $L = 1/\sqrt{1 + rD^2}$. This distribution allows analytical computation of spectral properties of $Z_j$. For example, the von Neumann entropy of $Z_j$ is $\mathcal{S}(Z_j) \approx -\mathrm{tr}[Z_j \log Z_j]$, which is given by

$$\mathcal{S}(Z_j) \approx -\frac{2}{\pi} \int_{-1}^{1} dt \sqrt{1 - t^2} (1 + DLt) \log\left(\frac{1}{D} + Lt\right).$$

This expression for large $D$ is approximately

$$\log D - \frac{2}{\pi} \int_{-1}^{1} dt \sqrt{1 - t^2} \left(1 + \frac{t}{\sqrt{r}}\right) \log\left(1 + \frac{t}{\sqrt{r}}\right),$$

and for large $D$ and $r$ the limiting expression is

$$\mathcal{S}(Z_j) \approx \log D - \frac{1}{8r} + O(r^{-2}). \tag{8}$$

The entropy deviates by at most an order one quantity from the maximal value $\log D$ that is attained for a maximally mixed state.

## IV. MATRIX PRODUCT STATES

MPS and their generalizations [34–37] provide efficient representations of quantum states by which classical simulation of quantum many-body systems becomes viable and are the natural representation of the density matrix renormalization group [38] and its tensor network generalizations. Applications range from efficient calculation of the ground state properties of quantum matter [39] to the outputs of quantum circuits [36]. They provide the tools for proving the existence of satisfying assignments in quantum satisfiability [40] and the area law [30]. Random MPS are the candidate boundary states for recent theoretical proposals of the theory of quantum gravity [3].

For the sake of concreteness, let us introduce the (one-dimensional) MPS on $2N + 1$ qudits, denoting $\vec{i} = \{i_{-N}, i_{-N+1} \ldots, i_{N-1}, i_N\}$:

$$|\psi(N)\rangle = \frac{1}{\mathcal{N}} \sum_{\vec{i}=1}^{d} \mathrm{tr}[A_{-N}^{i_{-N}} \ldots A_N^{i_N}] |i_{-N}, \ldots, i_N\rangle, \tag{9}$$

where $i_k$'s are the physical indices, $d$ is the local (physical) dimension of the Hilbert space, $A_k^{i_k}$'s are $D \times D \times d$ tensors, and $D$ is the bond dimension [34].

So far, rigorous results on generic MPS and their generalizations mainly focus on the translational-invariant case, where all the tensors in Eq. (9) are equal [31,41]. Other works focus on statistics of random MPS [42]. Here, we consider the general case in which translational invariance is relaxed. The most meaningful extensions pertaining to states of disordered systems and outputs of random quantum circuits require that the tensors are drawn only from a distribution. For example, if a Hamiltonian has local terms that are ergodic, then one expects the MPS representation of the states also to be ergodic. A similar statement is expected for the output state of a quantum circuit if the action of the circuit on the qubits is shift invariant.

Expectations in the MPS state $|\psi(N)\rangle$ can be related to a generalized quantum process $\phi_m(M) = \sum_{i=1}^{d} A_m^{i\dagger} M A_m^i$, $m \in \mathbb{Z}$. The linear maps $\phi_m$ are analogous to $\phi_i$ in Eq. (1), with $B^i = A_m^{i\dagger}$. However, they may not be quantum channels; although they are completely positive, they need not be trace preserving.

A preliminary observation is that the normalization constant $\mathcal{N}^2 = \sum_{\vec{i}} |\mathrm{tr}[A_{-N}^{i_N} \ldots A_N^{i_N}]|^2$ in Eq. (9) can be expressed as follows:

$$\mathcal{N}^2 = \sum_{\alpha,\beta=1}^{D} \langle\alpha|[\phi_N \circ \cdots \circ \phi_{-N}(|\alpha\rangle\langle\beta|)]|\beta\rangle,$$

where $|\alpha\rangle$, $\alpha = 1, \ldots, D$, are the elements of the computational basis on $\mathbb{C}^D$. The operators $|\alpha\rangle\langle\beta|$, $\alpha, \beta = 1, \ldots, D$, form a basis for the space of $D \times D$ matrices, allowing us to write the above identity as

$$\mathcal{N}^2 = \mathrm{Tr}[\phi_N \circ \cdots \circ \phi_{-N}], \tag{10}$$

where Tr denotes the trace of a superoperator, defined as $\mathrm{Tr}[\Phi] = \sum_{\alpha,\beta} \mathrm{tr}[|\beta\rangle\langle\alpha|\Phi(|\alpha\rangle\langle\beta|)]$.

There is a similar expression for the expectation of a local observable in the MPS state $|\psi(N)\rangle$. A local observable $O$ on the spins in $[m, n]$ is a linear operator on $\otimes_{j=m}^{n} \mathbb{C}^d$. Following the computation leading to Eq. (10), one may easily verify that the expectation of such an observable in the state $|\psi(N)\rangle$ is given by the expression

$$\langle\psi(N)|O|\psi(N)\rangle = \frac{\mathrm{Tr}[\phi_N \circ \cdots \circ \phi_{n+1} \circ \hat{O} \circ \phi_{m-1} \circ \cdots \circ \phi_{-N}]}{\mathrm{Tr}[\phi_N \circ \cdots \circ \phi_{-N}]}, \quad (11)$$

where $\hat{O}$ is the following superoperator:

$$\hat{O}(M) = \sum_{i,j=1}^{d} [\langle i_m, \ldots, i_n|O|j_m, \ldots, j_n\rangle \times A_n^{i_n \dagger} \ldots A_m^{i_m \dagger} M A_m^{j_m} \ldots A_n^{j_n}]. \quad (12)$$

In Theorem A.4 in Appendix A, we state and prove a generalization of Theorem 2 to non-trace-preserving quantum operations $\phi_m$ provided they satisfy $\phi_m(M) = 0$ for $M \geq 0$ if and only if $M = 0$ in addition to the above assumptions. The main new feature of the generalization is an ergodic analog of the left eigenvector of a channel, namely, a sequence $Z'_j$ satisfying

$$Z'_j = \frac{\phi_j^*(Z'_{j+1})}{\mathrm{tr}[\phi_j^*(Z'_{j+1})]}.$$

In the trace-invariant case, $Z'_j$ is the maximally mixed state for all $j$. Also, the shift equation (4) is replaced by $Z_j = \{\phi_j(Z_{j-1})/\mathrm{tr}[\phi_j(Z_{j-1})]\}$, ensuring that the matrices $Z_j$ have trace 1.

Using this result, we can directly calculate the thermodynamic limit of the expectation value of an observable with respect to an MPS [Eq. (9)]:

$$W(O) \equiv \lim_{N\to\infty} \langle\psi(N)|O|\psi(N)\rangle = \frac{\mathrm{tr}[Z'_{n+1}\hat{O}(Z_{m-1})]}{\mathrm{tr}[Z'_{n+1}\phi_n \circ \cdots \circ \phi_m(Z_{m-1})]}. \quad (13)$$

Furthermore, the entanglement spectrum of a bipartite partition of the infinite chain across the bond $j \sim j+1$, i.e., the spectrum of the reduced density matrix for one-half of the chain, corresponds to the eigenvalues of $R_j = Z_j Z'_{j+1}/\mathrm{tr}[Z_j Z'_{j+1}]$ (see Appendix C). For instance, the bipartite von Neumann entanglement entropy is

$$\mathcal{S}(j) = -\mathrm{tr}[R_j \log R_j]. \quad (14)$$

Since $Z_j$ and $Z'_{j+1}$ are positive, $R_j$ is similar to a positive matrix and $\log R_j$ is well defined.

Turning our attention to correlation functions, let $O_1$ and $O_2$ be local observables supported at (or near) the site $x = 0$. Let $O_1(x)$ and $O_2(x)$ denote the corresponding observables translated to have support at a general site $x$. We have the following.

*Theorem 2 [20].*—There exists $0 < \mu < 1$ such that, given local observables $O_1$ and $O_2$, the following correlation inequality holds:

$$|W[O_1(x)O_2(x+\ell)] - W[O_1(x)]W[O_2(x+\ell)]| \leq C\mu^{|\ell|}$$

with $C < \infty$ almost surely depending on $\mu$, $x$, $O_1$, and $O_2$.

In words, the two-point correlation function between local observables decays exponentially with the distance between their supports.

To the best of our knowledge, previous rigorous results considered only translationally invariant MPS. Equations (13) and (14) and Theorem 1 generalize these to the much larger class of MPS generated by ergodic sequences of matrices. In the translationally invariant case $\phi_0 = \phi_m$, the sequences $Z_n$ and $Z'_m$ are constant; i.e., $Z_{n+1} = Z_n$ and $Z'_m = Z'_{m+1}$ for all $n$, $m$. Equation (13) reduces to

$$W(O) = \frac{\mathrm{tr}[Z'_0\hat{O}(Z_0)]}{\mathrm{tr}[Z'_0\phi_0^{n-m}(Z_0)]},$$

the entropy $\mathcal{S}(j)$ is independent of $j$, and the correlation bound in Theorem 1 is independent of $x$.

## V. CONCLUDING REMARKS

In this paper, we present rigorous results on ergodic sequences of quantum channels with non-negligible decoherence. We show that, for an arbitrary initial density matrix, the state of the system subjected to such a sequence of channels converges exponential fast to an ergodic sequence of strictly positive density matrices. The ergodic framework of our results allows for noise models that are arbitrarily correlated.

In the second part, we prove results on the expectation values and two-point correlation functions of ergodic matrix product state. In particular, we rigorously show that the latter decays exponentially with the distance between the local observables.

Following the implications outlined in Sec. II A, we envision that these results could further pave the way for quantum computation by dissipation, shed light on the stability of nonequilibrium quantum phases of matter, and better characterize the output of random circuits in connection to near-term noisy quantum computers.

From a mathematical perspective, our result about MPS can be understood as describing correlations in the class of purely generated, ergodic, finitely correlated states, which generalizes the translation-invariant states considered in Ref. [43]. It would be of considerable interest to develop this theory further by extending it to ergodic, finitely correlated states over general $C^*$ algebras.

For these and other applications, it would be of technical interest to obtain asymptotics for the convergence factor $\mu$ and the prefactor $C$ in Theorems 2 and 1 with respect to the system size of channels and bond dimension of MPS, respectively. In the context of MPS, quantitative control on the convergence factor could make it possible to consider states with varying bond dimension as the system size increases.

## ACKNOWLEDGMENTS

We thank Frank Verstraete for helpful comments. R. M. acknowledges the support of the Frontiers institute and the support of MIT-IBM AI lab through the grant "Machine Learning in Hilbert Spaces." J. S. was supported by the National Science Foundation under Grants No. 1500386 and No. 1900015.

## APPENDIXES

We first lay out the basic notation and aspects of ergodic theory needed for our work. We then state the theorems. Although some of the proofs are to be found in Ref. [20], we outline the proof ideas. We then detail the calculations for independently and identically distributed random Haar channels and take the various asymptotic limits with respect to the system size and/or the environment. These prove Eqs. (5)–(10) of the main text. Lastly, the derivation of bipartite entanglement entropy [Eq. (17) in the main text] is given.

## APPENDIX A: FORMAL STATEMENT OF THE ERGODIC THEOREM

In this section, we state a result (Theorem A.3) that implies Theorem 1 in the main text. A sketch of the proof of Theorem A.3 is given. The full proof of this result is in Ref. [20].

To proceed, we need to introduce some basic notation and ideas. Let $\mathbb{M}_D$ denote the set of $D \times D$ complex matrices. We primarily use the trace norm on $\mathbb{M}_D$:

$$\|\mathrm{M}\|_1 = \mathrm{tr}[|M|].$$

Recall that $|M| = \sqrt{M^\dagger M}$, where $\sqrt{\cdot}$ denotes the operator square root. We also introduce the Hilbert-Schmidt inner product $\langle \tilde{M} M\rangle = \mathrm{tr}[\tilde{M}^\dagger M]$ and norm

$$\|\mathrm{M}\|_2^2 = \langle MM\rangle = \mathrm{tr}[M^\dagger M].$$

Recall that a *cone* in a vector space is a set that is closed under addition and multiplication by positive scalars. Let $\bar{E}$ denote the closed cone:

$$\bar{E} = \{\text{positive semidefinite}\, D \times D\, \text{matrices.}\}$$

The interior of $\bar{E}$ is

$$E = \{\text{positive definite}\, D \times D\, \text{matrices}\},$$

which is an open cone.

Let $\mathcal{L}(\mathbb{M}_D)$ denote the set of linear maps from $\mathbb{M}_D$ to $\mathbb{M}_D$. Given $\phi \in \mathcal{L}(\mathbb{M}_D)$, the adjoint $\phi^*$ is defined by

$$\mathrm{tr}[\tilde{M}^\dagger \phi(M)] = \mathrm{tr}[[\phi^*(\tilde{M})]^\dagger M].$$

A map $\phi \in \mathcal{L}(\mathbb{M}_D)$ is *positive* provided $\phi(M) \in \bar{E}$ whenever $M \in \bar{E}$. If, in addition, $\phi(M) \in E$ whenever $M \in \bar{E}$, then we say that $\phi$ is *strictly positive*. A *completely positive* map is one such that $\phi \otimes \mathbb{I}_{N\times N} : \mathbb{M}_D \otimes \mathbb{M}_N \to \mathbb{M}_D \otimes \mathbb{M}_N$ is positive for every $N$, where $\mathbb{I}_N$ denotes the identity map on $\mathbb{M}_N$. Kraus's theorem [1] states that $\phi$ is completely positive if and only if $\phi(M) = \sum_{i=1}^d B^i M B^{i\dagger}$ for some collection of matrices $B^i$, $i = 1, \ldots, d$. A map $\phi$ is *trace preserving* if $\mathrm{tr}[\phi(M)] = \mathrm{tr}[M]$ for all $M$; equivalently, $\phi^*(I) = I$. A *quantum channel* is a completely positive trace-preserving map.

Let $(\Omega, \mathcal{F}, \mathrm{Prob})$ be a probability space with $T : \Omega \to \Omega$ an invertible and ergodic map. Recall that $T$ is *ergodic* provided it is probability preserving and $\mathrm{Prob}[A] = 0$ or 1 for any event $A$ with $T^{-1}(A) = A$. Starting from a given completely positive map valued random variable, $\phi_0 : \Omega \to \{\text{completely positive maps}\}$, we consider the sequence of maps defined by evaluating $\phi_0$ along the trajectories of $T$:

$$\phi_{n;\omega} = \phi_{0;T^n\omega}. \tag{A1}$$

We follow the usual convention in probability theory and suppress the independent variable $\omega \in \Omega$ in most formulas; when it is needed—as in Eq. (A1)—we use a subscript to denote the value of a random variable at a particular $\omega \in \Omega$. Our main focus here is to study the action of the composition $\phi_n \circ \phi_{n-1} \circ \cdots \circ \phi_0$ of a long sequence of these maps. To obtain convergence, we require several assumptions. The first assumption is assumption 1 of the main text, restated here for the reader's convenience.

*Assumption A.1.*—For some $n_0 > 0$,

$$\mathrm{Prob}[\phi_{n_0} \circ \cdots \circ \phi_0 \text{ is strictly positive}] > 0.$$

*Remark.*—A map $\phi$ is strictly positive if and only if $\phi^*$ is strictly positive. Indeed, if $\phi$ is strictly positive and $M \in \bar{E}$, we have $\mathrm{tr}[\phi^*(M)M'] = \mathrm{tr}[M\phi(M')] > 0$ for any $M' \in \bar{E}$, as $\phi(M') > 0$. Thus, $\phi^*(M)$ is strictly positive. Thus, assumption A.1 implies that $\phi_0^* \circ \cdots \circ \phi_{n_0}^*$ is strictly positive with positive probability.

The second assumption, which for quantum channels is equivalent to assumption 2 of the main text, is the following.

*Assumption A.2.*—With probability one, $\ker \phi_0 \cap \bar{E} = \ker \phi_0^* \cap \bar{E} = \{0\}$. That is, if $\phi_0(M) = 0$ or $\phi_0^*(M) = 0$ with $M \in \bar{E}$, then $M = 0$.

*Remark.*—If $\phi_0$ is a quantum channel, then $\mathrm{tr}[\phi_0(M)] = \mathrm{tr}[M]$ for any $M$, so $\ker \phi_0 \cap \bar{E} = 0$. Thus, for quantum channels, assumption 2 is equivalent to the equality $\ker \phi_0^* \cap \bar{E} = \{0\}$ (assumption 2 of the main text). On the other hand, this identity is a nontrivial assumption for quantum channels. For example, if $\phi(M) = PMP + SMS^\dagger$ with $P$ a projection onto a subspace of dimension $D/2$ and $S$ a partial isometry from $\mathbb{I} - P$ to $P$, then $\phi$ is a channel but $\phi^*(\mathbb{I} - P) = 0$.

Given $n \in \mathbb{Z}$, let

$$\Phi_n = \begin{cases} \phi_n \circ \cdots \circ \phi_0 & \text{if } n > 0, \\ \phi_0 & \text{if } n = 0, \\ \phi_0 \circ \cdots \circ \phi_n & \text{if } n < 0. \end{cases}$$

By assumption A.1, the maps $\Phi_n$ preserve the cone $E$. In 1948, Krein and Rutman [44] derived a generalization of the classical Perron-Frobenius theorem to the context of linear maps preserving a convex cone. Evans and Høegh-Krohn [45] apply this result to obtain results for positive maps on $\mathcal{L}(\mathbb{M}_D)$. It follows from Theorem 2.3 in Ref. [45] that if $\Phi_n$ is strictly positive, then there is a unique (up to scaling) strictly positive matrix $R_n$ such that $\Phi_n(R_n) = \lambda_n R_n$, where $\lambda_n$ is the spectral radius of $\Phi_n$. Recall that the spectral radius is the maximum magnitude of all eigenvalues. Similarly, there is a unique (up to scaling) strictly positive matrix $L_n$ such that $\Phi_n^*(L_n) = \lambda_n L_n$. We normalize $R_n$ and $L_n$ so that $\mathrm{tr}[R_n] = \mathrm{tr}[L_n] = 1$.

Our main technical result is that $L_n$ converges as $n \to \infty$, while $R_n$ converges as $n \to -\infty$. This result generalizes a theorem of Hennion on the Perron-Frobenius eigenvectors of products of positive matrices [46].

*Theorem A.3 [20].*—The limits $\lim_{n\to-\infty} R_n = Z_0$ and $\lim_{n\to\infty} L_n = Z_0'$ exist almost surely and define random matrices $Z_0$ and $Z_0'$ that fall in $E$ with probability one. Furthermore, if we set $Z_n = Z_{0;T^n\omega}$ and $Z_n' = Z'_{0;T^n\omega}$, then

$$Z_n = \phi_n \cdot Z_{n-1} \quad \text{and} \quad Z_n' = \phi_n^* \cdot Z_{n+1}',$$

where $\cdot$ denotes the projective action of a positive map on the strictly positive $D \times D$ matrices of trace 1:

$$\phi_n \cdot M \equiv \frac{1}{\mathrm{tr}[\phi_n(M)]} \phi_n(M).$$

*Remark.*—If the maps $\phi_n$ are quantum channels, then $L_n = (1/D) I_D$, so $Z_n' = (1/D) I_D$ for all $n$.

Given $m < n$, let $P_{n,m}$ denote the rank-one operator

$$P_{n,m}(M) = \mathrm{tr}[Z_m' M] Z_n.$$

For $n - m$ large, the operator $\phi_n \circ \cdots \circ \phi_m$ is well approximated by $P_{n,m}$. To formulate this result precisely, we introduce the following norm for a linear map:

$$\|\Phi\|_1 = \max\{\mathrm{tr}[|\Phi(M)|] | \mathrm{tr}[|M|] = 1\}.$$

Theorem 2 of the main text is equivalent to the following.

*Theorem A.4 [20].*—Given $m < n$ in $\mathbb{Z}$, let $\Psi_{n,m} = \phi_n \circ \cdots \phi_m$. There is $0 < \mu < 1$ so that, given $x \in \mathbb{Z}$, the following bound holds:

$$\left\| \frac{1}{\mathrm{tr}[\Psi_{n,m}^*(I_D)]} \Psi_{n,m} - P_{n,m} \right\|_1 \le C_{\mu,x} \mu^{n-m}$$

for all $m \le x$ and $n \ge x$, with $C_{\mu,x} = C_{\mu,x;\omega}$ finite almost surely.

### 1. Sketch of the proofs

The key to proving Theorem A.3 is to show the existence of the limits $\lim_{n\to-\infty} R_n = Z_0$ and $\lim_{n\to\infty} L_n = Z_0'$. More generally, we show that

$$\lim_{n\to-\infty} \Phi_n \cdot Y_n \quad \text{and} \quad \lim_{n\to-\infty} \Phi_n^* \cdot Y_n \tag{A2}$$

exist for any sequence $Y_n$ of density matrices, and the limits are independent of the choice of the particular sequence. Indeed, once these limits are known, for any $n$ one takes $Z_n = Z_{0;T^n\omega}$ and $Z_n' = Z'_{0;T^n\omega}$ to obtain the full ergodic sequence. Since $\Phi_n \cdot R_n = R_n$, it follows that

$$Z_0 = \lim_{n\to-\infty} R_n = \lim_{n\to-\infty} \Phi_n \cdot R_n = \phi_0 \cdot (\lim_{n\to-\infty} \Phi_{n+1;T^{-1}\omega} \cdot R_n)$$
$$= \phi_0 \cdot Z_{-1}.$$

Similarly, $\Phi_n^* \cdot L_n = L_n$, and we find that

$$Z_0' = \lim_{n\to\infty} L_n = \lim_{n\to\infty} \Phi_n^* \cdot L_n = \phi_0^* \cdot (\lim_{n\to\infty} \Phi_{n-1;T^\omega} \cdot L_n) = \phi_0^* \cdot Z_1'.$$

The key to proving the existence of the limits in Eq. (A2) is to prove that $\Phi_n$ is exponentially contractive with respect to a suitable metric $d(Y_1, Y_2)$ on the space of density matrices:

$$\lim_{n\to\infty} \frac{1}{n} \log[d(\Phi_n \cdot Y_1, \Phi_n \cdot Y_2)] = \log \mu < 0. \tag{A3}$$

The existence of the limit in Eq. (A3) is guaranteed by Kingman's subadditve ergodic theorem [47]. A pivotal estimate in this context is the following:

$$d(\phi \cdot X, \phi \cdot Y) \le c(\phi) d(X, Y), \tag{A4}$$

where $c(\phi) \le 1$ is a quantity that depends on the particular completely positive map $\phi$. The subadditivity needed for the application of Kingman's theorem is provided by the inequality

$$c(\phi \circ \psi) \le c(\phi) c(\psi), \tag{A5}$$

which follows if we take $c(\phi)$ to be the smallest number such that Eq. (A4) holds. To obtain Eqs. (A4) and (A5) and to prove that the right-hand side of Eq. (A3) is less than zero, it is convenient to work with the specialized metric

$$d(Y_1, Y_2) = \frac{1 - m(Y_1, Y_2)m(Y_2, Y_1)}{1 + m(Y_1, Y_2)m(Y_2, Y_2)},$$

where $m(Y_1, Y_2) = \max\{\lambda \in \mathbb{R} | \lambda Y_2 \leq Y_1\}$. Although this metric is more convenient for the arguments, we show that it is equivalent to the trace-norm metric for strictly positive density matrices. Since the $\Phi_n$ are ultimately strictly positive by assumption 1 of the main text, the estimates obtained, in fact, hold in the trace norm. For details, see Ref. [20].

## APPENDIX B: INDEPENDENTLY AND IDENTICALLY DISTRIBUTED RANDOM HAAR CHANNELS

The results so far are purely existential; for a particular ergodic sequence of channels, one wishes to know the statistical properties of $Z_0$. To this end, we explicitly calculate them for two canonical examples. The first is for channels that are translation invariant which was detailed in the main text. The second class corresponds to the Kraus operators being IID Haar isometries (sections of independent Haar random unitary). The calculation of the latter case is the subject of this section.

Let $U_j$, $j \in \mathbb{Z}$, be an IID sequence of Haar distributed $L \times L$ unitary matrices, with $L = rD$, where one may think of $r$ and $D$ as the size of the environment and system, respectively. Consider the channels $\phi_j \in \mathcal{L}(\mathbb{M}_D)$ given by $\phi_j(M) = \mathrm{tr}_r[U_j(M \otimes Q_r)U_j^\dagger]$, where $Q_r$ is the $r \times r$ rank-one projector $Q_r \equiv |e_1\rangle\langle e_1|$ with $|e_1\rangle = (1, 0, \ldots, 0)^T$ and $\mathrm{tr}_r$ denotes the *partial trace* from $\mathbb{M}_L \to \mathbb{M}_D$. Note that the partial trace is the adjoint (with respect to the Hilbert-Schmidt inner product) of the map $M \mapsto M \otimes I_r$. The channel $\phi_j$ can be expressed in the Kraus form as

$$\phi_j(M) = \sum_{k=1}^{r} U_{j;k} M U_{j;k}^\dagger,$$

where the matrices $U_{j;k}$ are $D \times D$ blocks of $U_j$:

$$U_j = \begin{pmatrix} U_{j;1} & \cdots & \cdots \\ U_{j;2} & \ddots & \\ \vdots & & \ddots \\ U_{j;r} & \cdots & \cdots \end{pmatrix}.$$

It is easy to see that this particular sequence of channels satisfies assumptions 1 and 2. In fact, each $\phi_j$ is itself strictly positive with probability one. Thus, Theorem 1 applies, and the resulting stationary sequence $Z_j$, $j \in \mathbb{Z}$, satisfies the shift equation [Eq. (4) of the main text]

$$Z_j = \mathrm{tr}_r[U_j(Z_{j-1} \otimes Q_r)U_j^\dagger]. \tag{B1}$$

We study the distribution of $Z_j$ by the method of moments. As such, we are interested in computing the expectation $\mathbb{E}\{Z_j^{\otimes n}\}$, where $Z_j^{\otimes n} = Z_j \otimes \cdots \otimes Z_j$ with $n$ factors of $Z_j$. From $\mathbb{E}\{Z_j^{\otimes n}\}$, we can compute the expectation of any multilinear form $\prod_{\ell=1}^n \mathrm{tr}[M_\ell Z_j]$, via the identity:

$$\prod_{\ell=1}^{n} \mathrm{tr}[M_\ell Z_j] = \mathrm{tr}[(M_1 \otimes \cdots \otimes M_n) Z_j^{\otimes n}].$$

We begin computing the expectation for $n = 1$, which is the first moment $\mathbb{E}\{Z_j\}$. The key identity here is

$$\int_{U(L)} U(\alpha, \beta) U^\dagger(\beta', \alpha') dU = \frac{1}{L} \delta_\alpha^{\alpha'} \delta_\beta^{\beta'}. \tag{B2}$$

In terms of matrices, this identity can be expressed as

$$\int_{U(L)} U M U^\dagger dU = \mathrm{tr}[M] \frac{1}{L} I_L,$$

for an arbitrary $M \in \mathbb{C}^{L\times L}$. By Eq. (B1), the conditional expectation of $Z_j$ given $U_k$ for $k \neq j$ satisfies

$$\begin{aligned} \mathbb{E}\{Z_j | U_k, k \neq j\} &= \mathrm{tr}_r\left[\int_{U(L)} U(Z_{j-1} \otimes Q_r)U^\dagger dU\right] \\ &= \mathrm{tr}[Z_{j-1} \otimes Q_r] \frac{1}{L} \mathrm{tr}_r[I_L] = \frac{1}{D} I_D, \end{aligned}$$

since $\mathrm{tr}[Z_{j-1} \otimes Q_r] = \mathrm{tr}[Z_{j-1}] = 1$ and $\mathrm{tr}_r[I_L] = rI_D$. Averaging over all $U_k$ for $k \neq j$, we see that

$$\mathbb{E}\{Z_j\} = \frac{1}{D} I_D,$$

which is Eq. (7) in the main text. This result essentially proves that the expectation of $\phi_j$ is the completely depolarizing channel.

We turn now to the computation of $\mathbb{E}\{Z_j^{\otimes 2}\}$. The key is a generalization of Eq. (B2):

$$\begin{aligned} &\int_{U(L)} U(\alpha_1, \beta_1) U(\alpha_2, \beta_2) U^\dagger(\beta_1', \alpha_1') U^\dagger(\beta_2', \alpha_2') dU \\ &\quad = \mathrm{Wg}[(1,1), L](\delta_{\alpha_1}^{\alpha_1'} \delta_{\beta_1}^{\beta_1'} \delta_{\alpha_2}^{\alpha_2'} \delta_{\beta_2}^{\beta_2'} + \delta_{\alpha_1}^{\alpha_2'} \delta_{\beta_1}^{\beta_2'} \delta_{\alpha_2}^{\alpha_1'} \delta_{\beta_2}^{\beta_1'}) \\ &\qquad + \mathrm{Wg}[(2), L](\delta_{\alpha_1}^{\alpha_1'} \delta_{\beta_1}^{\beta_2'} \delta_{\alpha_2}^{\alpha_2'} \delta_{\beta_2}^{\beta_1'} + \delta_{\alpha_1}^{\alpha_2'} \delta_{\beta_1}^{\beta_1'} \delta_{\alpha_2}^{\alpha_1'} \delta_{\beta_2}^{\beta_2'}), \end{aligned} \tag{B3}$$

where $\mathrm{Wg}[(1,1), L] = (1/L^2 - 1)$ and $\mathrm{Wg}[(2), L] = -[1/L(L^2 - 1)]$ are *Weingarten functions* (see Ref. [33]). In terms of tensors, Eq. (B3) can be expressed as follows:

$$\int_{U(L)} (U \otimes U) M (U^\dagger \otimes U^\dagger) dU = \frac{1}{L^2-1} (\mathrm{tr}[M] E_L^{(1,1)} + \mathrm{tr}[E_L^{(2)} M] E_L^{(2)}) - \frac{1}{L(L^2-1)} (\mathrm{tr}[M] E_L^{(2)} + \mathrm{tr}[E_L^{(2)} M] E_L^{(1,1)}),$$

where $M$ is an arbitrary element of $\mathbb{M}_L \otimes \mathbb{M}_L$, $E_L^{(1,1)} = I_L \otimes I_L$ is the identity, and $E_L^{(2)}$ denotes the exchange, or SWAP, operator: $E_L^{(2)}(v \otimes w) = w \otimes v$.

Applying Eqs. (B1) and (B3) to $Z_j^{\otimes 2} = Z_j \otimes Z_j$, we find that

$$\mathbb{E}\{Z_j^{\otimes 2} | U_k, k \neq j\} = \frac{1}{L^2-1} \{\mathrm{tr}[(Z_{j-1} \otimes Q_r)^{\otimes 2}] \mathrm{tr}_r^{\otimes 2}[E_L^{(1,1)}] + \mathrm{tr}[E_L^{(2)} (Z_{j-1} \otimes Q_r)^{\otimes 2}] \mathrm{tr}_r^{\otimes 2}[E_L^{(2)}]\} - \frac{1}{L(L^2-1)} \{\mathrm{tr}[E_L^{(2)} (Z_{j-1} \otimes Q_r)^{\otimes 2}] \mathrm{tr}_r^{\otimes 2}[E_L^{(2)}] + \mathrm{tr}[(Z_{j-1} \otimes Q_r)^{\otimes 2}] \mathrm{tr}_r^{\otimes 2}[E_L^{(1,1)}]\},$$

where we introduce the notation

$$\mathrm{tr}_r^{\otimes 2}[M_1 \otimes M_2] = \mathrm{tr}_r[M_1] \otimes \mathrm{tr}_r[M_2],$$

extended by linearity to all of $\mathbb{M}_L \otimes \mathbb{M}_L$.

Now

$$\mathrm{tr}[(Z_{j-1} \otimes Q_r)^{\otimes 2}] = (\mathrm{tr}[Z_{j-1} \otimes Q_r])^2 = 1$$

and

$$\mathrm{tr}_r^{\otimes 2}[E_L^{(11)}] = \mathrm{tr}_r[I_L] \otimes \mathrm{tr}_r[I_L] = r^2 I_D \otimes I_D = r^2 E_D^{(1,1)}.$$

For the factors involving $E_L^{(2)}$, we use the identity

$$\mathrm{tr}[M_1 M_2] = \mathrm{tr}[E_L^{(2)} M_1 \otimes M_2]. \tag{B4}$$

Thus,

$$\mathrm{tr}[E_L^{(2)} (Z_{j-1} \otimes Q_r)^{\otimes 2}] = \mathrm{tr}[(Z_{j-1} \otimes Q_r)^2] = \mathrm{tr}[Z_{j-1}^2]$$

and $\mathrm{tr}_r^{\otimes 2}[E_L^{(2)}] = r E_D^{(2)}$. It follows that

$$\mathbb{E}\{Z_j^{\otimes 2} | U_k, k \neq j\} = \frac{1}{L(L^2-1)} \{(L - \mathrm{tr}[Z_{j-1}^2]) r^2 E_D^{(1,1)} + (L \mathrm{tr}[Z_{j-1}^2] - 1) r E_D^{(2)}\}.$$

Averaging over $U_k$, we obtain

$$\mathbb{E}\{Z_j^{\otimes 2}\} = \frac{1}{L(L^2-1)} [(L - m_{(2)}) r^2 E_D^{(1,1)} + (L m_{(2)} - 1) r E_D^{(2)}], \tag{B5}$$

where

$$m_{(2)} \equiv \mathbb{E}\{\mathrm{tr}[Z_{j-1}^2]|\} = \mathbb{E}\{\mathrm{tr}[Z_j^2]\},$$

since $Z_{j-1}$ and $Z_j$ are equidistributed.

To complete the calculation of $\mathbb{E}\{Z_j^{\otimes 2}\}$, we need to compute $m_{(2)}$. Note that $m_{(2)} = \mathbb{E}\{\mathrm{tr}[E_D^{(2)} Z_j^{\otimes 2}]\}$ by Eq. (B4). Since

$$E_D^{(2)} E_D^{(1,1)} = E_D^{(2)} \quad \text{and} \quad E_D^{(2)} E_D^{(2)} = E_D^{(1,1)},$$

it follows that

$$m_{(2)} = \frac{1}{L(L^2-1)} [(L - m_{(2)}) r^2 D + (L m_{(2)} - 1) r D^2].$$

Solving for $m_{(2)}$, we find that

$$m_{(2)} = \frac{(r+1)D}{rD^2+1}.$$

Plugging this result into Eq. (B5), we obtain

$$\mathbb{E}\{Z_j^{\otimes 2}\} = \frac{1}{rD^2+1} \left( r E_D^{(1,1)} + \frac{1}{D} E_D^{(2)} \right).$$

The variance of $Z_j$ can now be expressed as

$$\mathbb{E}\{Z_j^{\otimes 2}\} - \mathbb{E}\{Z_j\} \otimes \mathbb{E}\{Z_j\} = \frac{1}{rD^2+1} \left( \frac{1}{D} E_D^{(2)} - \frac{1}{D^2} E_D^{(1,1)} \right).$$

Note that $\mathrm{tr}[(1/D) E_D^{(2)}] = \mathrm{tr}[(1/D) E_D^{(1,1)}] = 1$. Thus, the prefactor $(rD^2+1)^{-1}$ sets the scale for the fluctuations around the mean $\mathbb{E}\{Z_j\} = (1/D) E_D^{(1,1)}$. To proceed, we introduce the notation $W_j$ for the rescaled deviation of $Z_j$ from the mean:

$$W_j = \sqrt{rD^2+1} \left( Z_j - \frac{1}{D} I_D \right). \tag{B6}$$

Thus,

$$\mathbb{E}\{W_j\} = 0 \quad \text{and} \quad \mathbb{E}\{W_j^{\otimes 2}\} = \frac{1}{D}E_D^{(2)} - \frac{1}{D^2}E_D^{(1,1)}. \tag{B7}$$

Our goal is to show that $W_j$ is asymptotically Gaussian, with covariance $(1/D)E_D^{(2)} - (1/D)E_D^{(1,1)}$. In other words, the distribution of $W_j$ is given by a traceless version of the Gaussian unitary ensemble [see Eq. (7)]. To obtain this result, we verify that $W_j$ asymptotically satisfies Wick's theorem.

Turning now to the computation of the $n$th moment for general $n$, we define $R^{(n)} = \mathbb{E}\{W_j^{\otimes n}\}$. By the shift invariance property, the right-hand side is independent of $j$, and using Eq. (B1) we see that

$$R^{(n)} = (rD^2+1)^{n/2} \times \mathbb{E}\left\{\left(\mathrm{tr}_r[U_j Z_{j-1} \otimes Q_r U_j^\dagger] - \frac{1}{D}I_D\right)^{\otimes n}\right\}. \tag{B8}$$

Because

$$\mathrm{tr}_r[U_j Z_{j-1} \otimes Q_r U_j^\dagger] - \frac{1}{D}I_D = \mathrm{tr}_r\left[U_j\left(Z_{j-1} \otimes Q_r - \frac{1}{L}I_L\right)U_j^\dagger\right],$$

we have

$$\left(\mathrm{tr}_r[U_j Z_{j-1} \otimes Q_r U_j^\dagger] - \frac{1}{D}I_D\right)^{\otimes n} = \mathrm{tr}_r^{\otimes n}\left[U_j^{\otimes n}\left(Z_{j-1} \otimes Q_r - \frac{1}{L}I_L\right)^{\otimes n} U_j^{\dagger\otimes n}\right].$$

There is a generalization of Eq. (B3) to higher-order products (see Ref. [33]). In terms of tensors, this generalization can be written as

$$\int_{U(L)} U^{\otimes n} M U^{\dagger\otimes n} dU = \sum_{\sigma,\tau\in S_n} \mathrm{Wg}(\tau\sigma, L)\mathrm{tr}[P_L^\tau M]P_L^\sigma, \tag{B9}$$

where Wg denotes the Weingarten function (see Ref. [33]), $S_n$ is the permutation group on $\{1,\ldots,n\}$, and for $v_k \in \mathbb{C}^L$ the operator $P_L^\alpha$ has the action

$$P_L^\alpha(v_1 \otimes\cdots\otimes v_j \otimes\cdots\otimes v_n) = v_{\alpha(1)} \otimes\cdots\otimes v_{\alpha(j)} \otimes\cdots v_{\alpha(n)}.$$

Applying Eq. (B9) to the average over $U_j$ in Eq. (B8) results in the following expression for $R^{(n)}$:

$$R^{(n)} = (rD^2+1)^{n/2}\sum_{\sigma\in S_n}\left(\sum_{\tau\in S_n}\mathrm{Wg}(\tau\sigma,L)\mathbb{E}\left\{\mathrm{tr}\left[P_L^\tau\left(Z_{j-1}\otimes Q_r - \frac{1}{L}I_L\right)^{\otimes n}\right]\right\}\right)\mathrm{tr}_r^{\otimes n}[P_L^\sigma]. \tag{B10}$$

The Weingarten function $\mathrm{Wg}(\alpha, L)$ depends only on the conjugacy class of the permutation $\alpha$: $\mathrm{Wg}(\tau\alpha\tau^{-1}, L) = \mathrm{Wg}(\alpha, L)$ (see Ref. [33]). Furthermore, $\mathrm{tr}[P_L^\tau M^{\otimes n}]$ depends only on the conjugacy class of $\tau$, as we now show. The conjugacy classes of $S_n$ are labeled by $m$-tuples $\vec{c} = (c_1, c_2\ldots, c_m)$ with $c_1 \geq c_2 \geq \cdots \geq c_m \geq 1$ and $\sum_j c_j = n$. A permutation in the class $(c_1, c_2\ldots, c_m)$ is given by

(1) a partition of $\{1,\ldots,n\}$ into $m$ sets of sizes $c_1,\ldots,c_m$ and
(2) a choice of a cyclic permutation on each subset of the partition.

A permutation is *cyclic* if it has no proper invariant subsets—there are $(j-1)!$ cyclic permutations in $S_j$. For any permutation $\tau \in S_n$ in the class $(c_1, c_2\ldots, c_m)$ and $M \in \mathbb{M}_L$, the trace $\mathrm{tr}[P_L^\tau M^{\otimes n}]$ is given by

$$\mathrm{tr}[P_L^\tau M^{\otimes n}] = \prod_{j=1}^m \mathrm{tr}[M^{c_j}]. \tag{B11}$$

Since $\mathrm{tr}[Z_{j-1} \otimes Q_r] = \mathrm{tr}[Z_{j-1}] = 1$, we see from Eq. (B11) that $\mathrm{tr}[P_L^\tau(Z_{j-1}\otimes Q_r - (1/L)I_L)^{\otimes n}] = 0$ for any permutation $\tau$ with a fixed point, in which case the conjugacy class has $c_m = 1$. In other words, the sum over $\tau$ in Eq. (B10) can be restricted to $\tau \in S_n'$, where

$$S_n' = \{\tau \in S_n | \tau(j) \neq j, j = 1,\ldots,n\}.$$

Let $CS_n$ denote the set of conjugacy classes of $S_n$. Given $\vec{c} \in CS_n$, let

$$E_L^{\vec{c}} = \sum_{\sigma\in\vec{c}} P_L^\sigma.$$

With the labeling of classes described above, this choice is consistent with the definitions of $E_L^{(1,1)}$ and $E_L^{(2)}$ given above. Associated to each class $\vec{c} = (c_1, c_2, \ldots, c_m) \in CS_n$, we define a distinguished permutation $[\vec{c}]$ which consists of the cycle

$$1 \mapsto 2 \mapsto \cdots \mapsto c_1 \mapsto 1$$

of the first $c_1$ numbers, followed by the cycle

$$c_1 + 1 \mapsto c_1 + 2 \mapsto \cdots \mapsto c_1 + c_2 \mapsto c_1 + 1$$

of the next $c_2$ numbers, etc., up to the last group of $c_m$ numbers:

$$\sum_{j=1}^{m-1} c_j + 1 \mapsto \sum_{j=1}^{m-1} c_j + 2 \mapsto \cdots \mapsto n \mapsto \sum_{j=1}^{m-1} c_j.$$

That is, $[\vec{c}] = (1,2,\ldots,c_1)(c_1+1,\ldots,c_1+c_2)\cdots(c_1+\cdots+c_{m-1},\ldots,n)$.

Putting all of this together, we obtain the following formula:

$$R^{(n)} = \mathbb{E}\{W_j^{\otimes n}\} = \sum_{\vec{c}\in CS_n} w(\vec{c};r,D)\frac{1}{D^{m(\vec{c})}}E_D^{\vec{c}}, \qquad \text{(B12)}$$

where $m(\vec{c})$ denotes the number of cycles in the conjugacy class $\vec{c}$, i.e., $m(\vec{c}) = m$ if $\vec{c} = (c_1,\ldots,c_m)$ with $c_m \geq 1$,

$$w(\vec{c};r,D) = L^{m(\vec{c})}(rD^2+1)^{n/2}\sum_{\tau\in S_n'}\left[\mathrm{Wg}(\tau[\vec{c}],L) \times \mathbb{E}\left\{\mathrm{tr}\left[P_L^\tau\left(Z_{j-1}\otimes Q_r - \frac{1}{L}I_L\right)^{\otimes n}\right]\right\}\right], \qquad \text{(B13)}$$

and we observe that $\mathrm{tr}_r^{\otimes n}[E_L^{\vec{c}}] = r^{m(\vec{c})}E_D^{\vec{c}}$. Adding and subtracting $(1/D)I_D\otimes Q_r = \mathbb{E}\{Z_{j-1}\}\otimes Q_r$ inside the parentheses, we obtain the following alternative expression for $w(c;\vec{r},s)$:

$$w(\vec{c};r,D) = L^{m(\vec{c})}\sum_{\tau\in S_n'}\mathrm{Wg}(\tau[\vec{c}],L)\mathbb{E}\left\{\mathrm{tr}\left[P_L^\tau\left(W_{j-1}\otimes Q_r + \sqrt{rD^2+1}\left(\frac{1}{D}I_D\otimes Q_r - \frac{1}{L}I_L\right)\right)^{\otimes n}\right]\right\}. \qquad \text{(B14)}$$

Let $\tau\in S_n'$ and let

$$E(\tau) = \mathbb{E}\left\{\mathrm{tr}\left[P_L^\tau\left(W_{j-1}\otimes Q_r + \sqrt{rD^2+1}\left(\frac{1}{D}I_D\otimes Q_r - \frac{1}{L}I_L\right)\right)^{\otimes n}\right]\right\},$$

which is the expectation appearing in the sum on the right-hand side in Eq. (B14). The conjugacy class of $\tau$, $[\tau] = (c_1'c_2'\ldots c_m')$, satisfies $c_m'\geq 2$. Given commuting matrices $A$ and $B$,

$$\mathrm{tr}[P_L^\tau(A+B)^{\otimes n}] = \prod_{\ell=1}^m\sum_{k_\ell=0}^{c_\ell'}\binom{c_\ell'}{k_\ell}\mathrm{tr}[A^{k_\ell}B^{c_\ell'-k_\ell}].$$

Applying this expression to $E(\tau)$, we obtain the following expression for $E(\tau)$:

$$\mathbb{E}\left\{\prod_{\ell=1}^m\left[(1-(1-r)^{1-c_\ell'})D\left(\frac{(r-1)\sqrt{rD^2+1}}{L}\right)^{c_\ell'} + \sum_{k_\ell=2}^{c_\ell'}\binom{c_\ell'}{k_\ell}\left(\frac{(r-1)\sqrt{rD^2+1}}{L}\right)^{c_\ell'-k_\ell}\mathrm{tr}[W_{j-1}^{k_\ell}]\right]\right\},$$

where we observe that

$$\mathrm{tr}\left[\left(\frac{1}{D}I_D\otimes Q_r - \frac{1}{L}I_L\right)^c\right] = \left(\frac{r-1}{L}\right)^c[1-(1-r)^{1-c}]D$$

and

$$(W_{j-1}\otimes Q_r)\left(\frac{1}{D}I_D\otimes Q_r - \frac{1}{L}I_L\right) = \frac{r-1}{L}W_{j-1}\otimes Q_r.$$

On the right-hand side, the terms with $k_\ell = 1$ are excluded, as they vanish since $\mathrm{tr}[W_{j-1}] = 0$.

Given $m$-tuples of non-negative integers $\vec{k}$ and $\vec{c}'$, let $\vec{k}\prec\vec{c}'$ denote the relation $k_\ell\leq c_\ell'$ for all $\ell$. Let $[\vec{k}]$ denote the permutation corresponding to a $k_1$-cycle of the first $k_1$ integers, a $k_2$-cycle of the next $k_2$, etc., up to the last $k_m$ integers. Zero cycles are dropped, so $[\vec{k}]$ is an element of $S_{|\vec{k}|}$, where $|\vec{k}| = \sum_\ell k_\ell$. Note that

$$\prod_{\ell=1}^m\mathrm{tr}[W_{j-1}^{k_\ell}] = \mathrm{tr}[P_D^{[\vec{k}]}W_{j-1}^{(|\vec{k}|)}].$$

Thus, for $\vec{c}'\in CS_n'$ and $\tau\in\vec{c}'$,

$$E(\tau) = \sum_{\vec{k}\prec c' k_\ell\neq 1}\left[\binom{\vec{c}'}{\vec{k}}\left(\frac{(r-1)\sqrt{rD^2+1}}{rD}\right)^{n-|\vec{k}|}D^{\#\{k_\ell=0\}} \times\prod_{\ell:k_\ell=0}\{1-(1-r)^{1-c_\ell'}\}\mathrm{tr}[P_D^{[\vec{k}]}R^{(|\vec{k}|)}]\right],$$

where $\binom{\vec{c}'}{\vec{k}} = \prod_{\ell=1}^m\binom{c_\ell'}{k_\ell}$.

Returning to the weights for Eq. (B12), we find that

$$w(\vec{c};r,D) = (rD)^{m(\vec{c})}\sum_{\tau\in S'_n}\Bigg\{\mathrm{Wg}(\tau[\vec{c}],rD) \times \sum_{\substack{\vec{k}<[\tau]\\ k_\ell\neq 1}}\Bigg[\binom{[\tau]}{\vec{k}}\left(\frac{\sqrt{1+rD^2}(r-1)}{rD}\right)^{n-|\vec{k}|}D^{\#\{k_\ell=0\}} \times \prod_{\ell:k_\ell=0}[1-(1-r)^{1-c'_\ell(\tau)}]\mathrm{tr}[P_D^{[\vec{k}]}R^{(|\vec{k}|)}]\Bigg]\Bigg\}. \tag{B15}$$

Note that $\mathrm{tr}[P_s^{[\vec{k}]}R^{(|\vec{k}|)}]$ depends only on the nonzero elements of $\vec{k}$ and is unchanged if we permute the entries of $\vec{k}$.

Given $\vec{c}\in CS'_n$, let $M(\vec{c};r,D)=\mathrm{tr}[P_D^{[\vec{c}]}R^{(n)}]$. Using Eq. (B12) to compute $M(\vec{c};r,D)$, we find

$$M(\vec{c};r,D) = \sum_{\sigma\in S_n} w([\sigma];r,D)\frac{D^{m([\vec{c}]\sigma)}}{D^{m(\sigma)}}.$$

Here, $m(\sigma)=m([\sigma])$ denotes the number of cycles in the permutation $\sigma$. Using Eq. (B12), we obtain the following linear equation for the moments $M(\vec{c};r,D)$:

$$M(\vec{c};r,D) = \sum_{\sigma\in S_n}\sum_{\tau\in S'_n}\sum_{\substack{\vec{k}<[\tau]\\ k_\ell\neq 1}} M(\vec{k};r,D)\Bigg\{\mathrm{Wg}(\tau\sigma,rD)r^{m(\sigma)}D^{m([\vec{c}]\sigma)}\binom{[\tau]}{\vec{k}}\left(\frac{\sqrt{1+rD^2}(r-1)}{rD}\right)^{n-|\vec{k}|}D^{\#\{k_\ell=0\}} \times \prod_{\ell:k_\ell=0}[1-(1-r)^{1-c'_\ell(\tau)}]\Bigg\}. \tag{B16}$$

Here, $M(\vec{k};r,s) \coloneqq M(\vec{k}';r,s)$, where $\vec{k}'$ consists of the nonzero elements of $\vec{k}$ listed in decreasing order. In principle, Eq. (B16) allows us to solve for all moments $M(\vec{c};r,s)$. Note, in particular, the triangular structure to the equations: to solve for the moments with $\vec{c}\in CS_n$, we need to know only the moments of lower order. Thus, the equations can be solved recursively for increasing $n$.

To analyze the solution to Eq. (B16) in the limits $r\to\infty$ and $D\to\infty$, we need the Weingarten function asymptotics (see Ref. [33]). Given $\alpha\in S_n$, with $[\alpha]=(c_1c_2\ldots c_m)$,

$$\mathrm{Wg}(\alpha,L) = C_\alpha L^{-n-|\alpha|}(-1)^{|\alpha|}+O(L^{-n-|\alpha|-2}), \tag{B17}$$

where $|\alpha|=n-m$, which is the minimal length of $\alpha$ as a product of 2-cycles, and

$$C_\alpha = \prod_{i=1}^m\frac{(2c_i)!}{c_i!(c_i+1)!}.$$

Note that $d(\alpha,\beta)=|\alpha\beta^{-1}|$ is a metric on $S_n$. In particular, we have the triangle inequality

$$|\alpha\gamma^{-1}| \leq |\alpha\beta^{-1}|+|\beta\gamma^{-1}|. \tag{B18}$$

Furthermore, $|\alpha|$ is invariant on conjugacy classes; in particular, $|\alpha|=|\alpha^{-1}|$.

### 1. Gaussianity: Limit of large *r*

We now consider the asymptotics of Eq. (B16) as $r\to\infty$. This example is technically somewhat simpler than the large $D$ limit. Let us denote by $A(r)\sim B(r)$ the asymptotic relation $\lim_{r\to\infty}A(r)/B(r)=1$. By Eq. (B17),

$$\mathrm{Wg}(\tau\sigma,rD)r^{m(\sigma)}\left(\frac{\sqrt{1+rD^2}(r-1)}{sD}\right)^{n-|\vec{k}|} \sim (-1)^{|\tau\sigma|}C_{\tau\sigma}D^{-n-|\tau\sigma|}\frac{r^{(n-|\vec{k}|)/2}}{r^{|\sigma|+|\tau\sigma|}}.$$

By the triangle inequality [Eq. (B18)], $|\sigma|+|\tau\sigma|\geq|\tau|$. Furthermore, for any $\tau\in S'_n$, we have $|\tau|\geq n/2$. Therefore, unless $|\vec{k}|=0$ and $|\sigma|+|\sigma\tau|=|\tau|=n/2$, it follows that

$$\mathrm{Wg}(\tau\sigma,rD)r^{m(\sigma)}\left(\frac{\sqrt{1+rD^2}(r-1)}{Dr}\right)^{n-|\vec{k}|}\to 0$$

as $r\to\infty$. The identity $|\sigma|+|\sigma\tau|=n/2$ holds if and only if $\tau$ is a pairing (i.e., a product of $n/2$ disjoint 2-cycles) and if $\sigma$ is a product of disjoint 2-cycles contained in $\tau$ (denoted $\sigma\prec\tau$). In this case, we have $C_{\tau\sigma}=1$ and

$$\mathrm{Wg}(\tau\sigma,rD)r^{m(\sigma)}\left(\frac{\sqrt{1+rD^2}(r-1)}{Dr}\right)^{n-|\vec{k}|} \to (-1)^{|\tau\sigma|}D^{-n-|\tau\sigma|}.$$

Since all terms on the right-hand side in Eq. (B16) tend to zero except for those with $\vec{k}=0$ and $\tau$ is a pairing, this equation reduces to

$$M(\vec{c};r\to\infty,D) = \sum_{\tau\in P_n\sigma\prec\tau}(-1)^{|\tau\sigma|}D^{(n/2)-|[\vec{c}]\sigma|-|\tau\sigma|},$$

where $P_n$ denotes the set of pairings of $\{1,\ldots,n\}$. It follows that all moments are of the order of one as $r\to\infty$ and given by the above expression. In particular, $M(\vec{c}; r \to \infty, D)$ vanishes if $\vec{c} \in CS_n$ for $n$ odd.

Furthermore, we see that $w(\vec{c}; r \to \infty, D) = 0$ unless $n$ is even and

$$\vec{c} = (\underbrace{2, \ldots, 2}_{j}, \underbrace{1, \ldots, 1}_{n-2j}) \equiv (2^j, 1^{n-2j})$$

for some $j$, for which we have

$$w[(2^j, 1^{n-2j}); r \to \infty, D] = (-1)^{(n/2)-j} \frac{(n-2j)!}{2^{(n/2)-j}(\frac{n}{2}-j)!},$$

where $[(n-2j)!/2^{(n/2)-j}(\frac{n}{2}-j)!]$ is the number of pairings of $\{1, \ldots, n-2j\}$. Thus,

$$\lim_{r\to\infty} R^{(n)} = \begin{cases} 0 & \text{if } n \text{ is odd,} \\ \sum_{j=0}^{n/2} (-1)^{(n/2)-j} \frac{(n-2j)!}{2^{(n/2)-j}(\frac{n}{2}-j)!} \frac{1}{D^{n-j}} E_D^{(2^j, 1^{n-2j})} & \text{if } n \text{ is even.} \end{cases} \tag{B19}$$

This result reduces to Eq. (B7) for $n = 1$ and $n = 2$. For higher $n$, the result it gives is the same as Wick's theorem for $\lim_r \mathbb{E}\{W_j^{\otimes n}\}$, showing that these matrices have a Gaussian distribution in the large $r$ limit. For example,

$$\lim_{r\to\infty} R^{(4)} = \frac{1}{D^2} E_D^{(2,2)} - \frac{1}{D^3} E_D^{(2,1,1)} + \frac{3}{D^4} E^{(1,1,1,1)}.$$

In more detail, note that Wick's theorem would imply that $\lim_r \mathbb{E}\{W_j^{\otimes 2n}\}$ is a sum of terms obtained by pairing the matrices in the tensor product and averaging according to the variance in Eq. (B7), namely, $R^{(2)} = (1/D)E_D^{(2)} - (1/D^2)E_D^{(1,1)}$. Further expanding each term in the sum according to a choice of one of the two terms in $R^{(2)}$ for each pair of factors results in an expression for $\mathbb{E}\{W_j^{\otimes 2n}\}$ as a sum over permutations with only two-cycles and one-cycles. In this expansion, each permutation comes with the weight

$$\#\text{pairings of the set covered by one cycles} \\ \times (-1)^{(1/2)\#\text{ one-cycles}} \frac{1}{D^{\#\text{cycles}}}.$$

The first term counts the number of ways that a given permutation can arise in the above expansion, while the second term comes from the weights attached to $E_D^{(2)}$ and $E_D^{(1,1)}$ in the variance. As one may easily check, the result is identical with the right-hand side of Eq. (B19).

### 2. Gaussianity: Large $D$ limit

For the large $D$ limit, we should first examine what happens to the moments like $M[(2); r, D]$ as $D \to \infty$. Note that $M[(2); r, D] = \mathrm{tr}[P_D^{[(2)]} R^{(2)}] = D - (1/D)$. Thus, we expect moments to become large, and a further normalization is needed to compute the limit of Eq. (B16). Indeed, we have

$$\mathrm{Wg}(\tau\sigma, rD) D^{m([\vec{c}]\sigma)} D^{\#\{k_\ell = 0\}} \left(\frac{\sqrt{1 + rD^2}(r-1)}{Dr}\right)^{n-|\vec{k}|} \sim (-1)^{|\tau\sigma|} C_{\tau\sigma} \frac{r^{(n-|\vec{k}|)/2}}{r^{|\sigma|+|\tau\sigma|}} D^{\#\{k_\ell=0\} - |[\vec{c}]\sigma| - |\tau\sigma|}.$$

By the triangle inequality, $|[\vec{c}]\sigma| + |\tau\sigma| \geq |[\vec{c}]\tau^{-1}| \geq 0$, with equality precisely if $[\vec{c}] = \tau$ and $\sigma \prec \tau$. Everything is maximized for $\vec{k} = 0$ and $\tau = [\vec{c}]$, suggesting that

$$M[(2^j); r, D] \sim \mathrm{const} D^{m(\vec{c})}.$$

To proceed, we define

$$\tilde{M}(\vec{c}; r, D) = D^{-m(\vec{c})} M(\vec{c}; r, D),$$

so $\tilde{M}[(2); r, D \to \infty] = 1$. Working from Eq. (B16), we find that

$$\tilde{M}(\vec{c}; r, D) = \sum_{\sigma \in S_n} \sum_{\tau \in S'_n} \sum_{\substack{\vec{k} \prec [\tau] \\ k_\ell \neq 1}} \tilde{M}(\vec{k}; r, D) \Bigg\{ \mathrm{Wg}(\tau\sigma, rs) r^{m(\sigma)} D^{m([\vec{c}]\sigma) - m(\vec{c}) + m(\tau)} \binom{[\tau]}{\vec{k}} \left(\frac{\sqrt{1 + rD^2}(r-1)}{Dr}\right)^{n-|\vec{k}|} \\ \times \prod_{\ell : k_\ell = 0} [1 - (1-r)^{1-c'_\ell(\tau)}] \Bigg\}.$$

The large $D$ asymptotics of the coefficients follow from

$$\mathrm{Wg}(\tau\sigma, rD) r^{m(\sigma)} D^{m([\vec{c}]\sigma)-m(\vec{c})+m(\tau)} \sim (-1)^{|\tau\sigma|} C_{\tau\sigma} \frac{r^{(n-|\vec{k}|)/2}}{r^{|\sigma|+|\tau\sigma|}} D^{|[\vec{c}]|-|\tau|-|[\vec{c}]\sigma|-|\tau\sigma|}.$$

Again by the triangle inequality, $|\tau| + |[\vec{c}]\sigma| + |\tau\sigma| \geq |\tau| + |[\vec{c}]\tau^{-1}| \geq |[\vec{c}]|$, so the limiting equation for $\tilde{M}(\vec{c}; r, s)$ has order one coefficients as $D \to \infty$. Thus, the resulting solution is of the order of one.

Rewriting Eq. (B15) for the weights $w(\vec{c}; r, D)$ in terms of $\tilde{M}$, we find that

$$w(\vec{c}; r, D) = \sum_{\tau \in S'_n} \sum_{\substack{\vec{k} \prec [\tau] \\ k_\ell \neq 1}} \tilde{M}(\vec{k}; r, D) \left\{ \mathrm{Wg}(\tau[\vec{c}], rD)(rD)^{m(\vec{c})} D^{m(\tau)} \binom{[\tau]}{\vec{k}} \left( \frac{\sqrt{1 + rD^2(r-1)}}{rD} \right)^{n-|\vec{k}|} \prod_{\ell : k_\ell = 0} [1 - (1-r)^{1-c'_\ell(\tau)}] \right\}.$$

By Eq. (B17), the first three factors of the coefficient of $\tilde{M}(\vec{k}; r, D)$ are asymptotically

$$\mathrm{Wg}(\tau[\vec{c}], rD)(rD)^{m(\vec{c})} D^{m(\tau)} \sim (-1)^{|\tau[\vec{c}]|} C_{\tau[\vec{c}]} \frac{r^{(n-|\vec{k}|)/2}}{r^{|[\vec{c}]|+|\tau\sigma|}} D^{n-|\tau|-|[\vec{c}]|-|\tau[\vec{c}]|}.$$

Now $|\tau| + |[\vec{c}]| + |\tau[\vec{c}]| \geq 2|\tau| \geq n$, with equality if and only if $\tau$ is a pairing and $[\vec{c}] \prec \tau$. Thus, $w(\vec{c}; r, D \to \infty) = 0$ unless $n$ is even and $\vec{c} = (2^j, 1^{n-2j})$ for some $j$; for such $\vec{c}$, we have

$$w[(2^j, 1^{n-2j}); r, D \to \infty] = (-1)^{n-2j} p_{n-2j} \sum_{\substack{\vec{k} \prec (2^{\frac{n}{2}}) \\ k_\ell \neq 1}} r^{-|\vec{k}|/2} \left( \frac{r-1}{r} \right)^{\#\{k_\ell = 0\}} \tilde{M}(\vec{k}; r, D \to \infty).$$

Thus,

$$\tilde{M}[(2^m); r, D \to \infty] \sim \sum_{j=0}^{m} \sum_{\sigma \in (2^j 1^{2m-2j})} (-1)^{2m-2j} p_{2m-2j} \sum_{\substack{\vec{k} \prec (2^m) \\ k_\ell \neq 1}} \left[ \tilde{M}(\vec{k}; r, D \to \infty) r^{-|\vec{k}|/2} \left( \frac{r-1}{r} \right)^{\#\{k_\ell = 0\}} \frac{1}{D^{3m-j}} \mathrm{tr}[P_D^{(2^m)} P_D^{\sigma}] \right].$$

The trace $\mathrm{tr}[P_D^{(2^m)} P_D^{\sigma}]$ is bounded by $D^{m+j}$, with equality precisely if $\sigma \prec [(2^m)]$. As a result, the only terms that survive in the limit correspond to $j = m$ and $\sigma = [(2^m)]$. Thus,

$$\tilde{M}[(2^m); r, D \to \infty] = \sum_{\substack{\vec{k} \prec (2^m) \\ k_\ell \neq 1}} r^{-|\vec{k}|/2} \left( \frac{r-1}{r} \right)^{\#\{k_\ell = 0\}} \tilde{M}(\vec{k}; r, D \to \infty).$$

Introducing $q(m, r) = \tilde{M}[(2^m); r, D \to \infty]$, this equation reduces to

$$q(m, r) = \sum_{j=0}^{m} \binom{m}{j} \left( \frac{r-1}{r} \right)^{m-j} r^{-j} q(j, r) = \frac{1}{r^m} \sum_{j=0}^{m} \binom{m}{j} (r-1)^{m-j} q(j, r).$$

We have $q(0, r) = 1$ (by definition). It follows that $q(m, r) = 1$ for all $m$ by induction. Thus,

$$w[(2^j, 1^{n-2j}); r, D \to \infty] = (-1)^{n-2j} \frac{(n-2j)!}{2^{(n/2)-j}(\frac{n}{2} - j)!} \sum_{\substack{\vec{k} \prec (2^{n/2}) \\ k_\ell \neq 1}} r^{-|\vec{k}|/2} \left( \frac{r-1}{r} \right)^{\#\{k_\ell = 0\}} = (-1)^{n-2j} \frac{(n-2j)!}{2^{(n/2)-j}(\frac{n}{2} - j)!},$$

which proves that the limit is Gaussian in this case.

## APPENDIX C: BIPARTITE ENTANGLEMENT ENTROPY OF AN ERGODIC MATRIX PRODUCT STATE

In this section, we prove the formula for the bipartite entanglement entropy for ergodic MPS given by Eq. (17) in the main text. By translation invariance in distribution, it suffices to consider $j = 0$. To focus on the bipartite entanglement entropy across a single cut, we make use of open boundary conditions for the chain.

*Remark C.1.*—The thermodynamic limit of the chain with open boundary conditions is identical to that with periodic boundary conditions; this limit follows from the convergence given in Theorem 2 of the main text.

Consider an MPS defined over $[-M, N]$, with $M, N > 0$ and open boundary conditions. For this state, Eq. (11) of the main text becomes

$$|\psi(-M,N)\rangle = \sum_{\vec{i}=1}^{d}[\vec{b}_{-M}^{i_{-M};\dagger}A_{-M+1}^{i_{-M+1}}\dots A_{N-1}^{i_{N-1}}\vec{b}_N^{i_N} \times |i_{-M},\dots,i_N\rangle], \quad \text{(C1)}$$

where $\vec{b}^i_{-M}$ and $\vec{b}^i_N$, $i = 1, \dots, d$, are nonzero vectors in $\mathbb{C}^d$ specifying the boundary conditions at each end of the chain.

We consider the thermodynamic limit in two steps: first $N \to \infty$ and then $M \to \infty$. For an observable $O$ on the spins in $[-M, 0]$, it follows from Eq. (C1) that the expectation of $O$ is given by the following modification of Eq. (12) of the main text:

$$\langle\psi(-M,N)O\psi(-M,N)\rangle = \frac{\mathrm{tr}[B_N\phi_{N-1}\circ\cdots\phi_1(\hat{O}_{-M})]}{\mathrm{tr}[B_N\phi_{N-1}\cdots\circ\cdots\phi_{-M+1}(B_{-M})]}, \quad \text{(C2)}$$

where

$$B_N = \sum_{i=1}^{d}\vec{b}_N^i\vec{b}_N^{i;\dagger}, \qquad B_{-M} = \sum_{i=1}^{d}\vec{b}_{-M}^i\vec{b}_{-M}^{i;\dagger},$$

and

$$\hat{O}_{-M} = \sum_{\vec{i},\vec{j}=1}^{d}[\langle i_{-M},\dots,i_0|O|j_{-M},\dots,j_0\rangle \times A_0^{i_0\dagger}\dots A_{-M+1}^{i_{-M+1}\dagger}\vec{b}_{-M}^{i_{-M}}\vec{b}_{-M}^{j_{-M}\dagger}A_{-M+1}^{j_{-M+1}}\dots A_0^{j_0}].$$

Note that in the present case $\hat{O}_{-M}$ is a $D \times D$ matrix.

It follows from Eq. (C2) and Theorem 2 of the main text that

$$\lim_{N\to\infty}\langle\psi(-M,N)|O|\psi(-M,N)\rangle = \frac{\mathrm{tr}[Z_1'\hat{O}_{-M}]}{\mathrm{tr}[Z_1'\phi_0\circ\cdots\phi_{-M+1}(B_{-M})]}. \quad \text{(C3)}$$

Let $\rho_{-M}$ denote the reduced density matrix of $[-M, 0]$, in the limit $N \to \infty$. By Eq. (C3), we have

$$\langle j_{-M},\dots,j_0|\rho_{-M}|i_{-M},\dots,i_0\rangle = \frac{\mathrm{tr}[Z_1'A_0^{i_0\dagger}\dots A_{-M+1}^{i_{-M+1}\dagger}\vec{b}_{-M}^{i_{-M}}\vec{b}_{-M}^{j_{-M}\dagger}A_{-M+1}^{j_{-M+1}}\dots A_0^{j_0}]}{\mathrm{tr}[Z_1'\phi_0\circ\cdots\phi_{-M+1}(B_{-M})]}.$$

Let

$$Q_{-M} \equiv \sqrt{Z_1'}\{\phi_0\circ\cdots\circ\phi_{-M+1}(B_{-M})\}\sqrt{Z_1'}.$$

Note that $Q_{-M}$ is a positive matrix. Let $\vec{v}_1, \dots, \vec{v}_D$ be its orthonormal eigenvectors, with corresponding eigenvalues $\lambda_1 \geq \cdots \geq \lambda_D \geq 0$. Let

$$|\alpha_j\rangle = \frac{1}{\sqrt{\lambda_j}}\sum_{\vec{i}=1}^{d}\vec{b}_{-M}^{i_{-M}\dagger}A_{-M+1}^{i_{-M+1}}\dots A_0^{i_0}\sqrt{Z_1'}\vec{v}_j.$$

Then

$$\rho_{-M} = \frac{\sum_{j=1}^{D}\lambda_j|\alpha_j\rangle\langle\alpha_j|}{\mathrm{tr}[Q_{-M}]} = \frac{\sum_{j=1}^{D}\lambda_j|\alpha_j\rangle\langle\alpha_j|}{\sum_{j=1}^{D}\lambda_j}$$

and

$$\langle\alpha_j|\alpha_k\rangle = \frac{1}{\sqrt{\lambda_j\lambda_k}}\vec{v}_j^\dagger Q_{-M}\vec{v}_k = \delta_j^k.$$

Thus, $|\alpha\rangle_j$ are eigenvectors of $\rho_{-M}$, and the nonzero eigenvalues of $\rho_{-M}$ are given by $\lambda_i/\sum_j\lambda_j$.

It follows that the bipartite entropy across the bond $0 \sim 1$ (in the limit $N \to \infty$ with $M < \infty$) is given by

$$S_{-M}(0) = -\mathrm{tr}[\tilde{Q}_{-M}\log\tilde{Q}_{-M}],$$

with $\tilde{Q}_{-M} = (1/\mathrm{tr}[Q_{-M}])Q_{-M}$. Equation (17) of the main text follows, since

$$\lim_{M\to\infty}\tilde{Q}_{-M} = \frac{1}{\mathrm{tr}[\sqrt{Z_1'}Z_0\sqrt{Z_1'}]}\sqrt{Z_1'}Z_0\sqrt{Z_1'}$$

by Theorem 2 of the main text.